\documentclass[10pt]{article}
\usepackage[preprint,nonatbib]{neurips_2025}
\usepackage[T1]{fontenc}
\usepackage[utf8]{inputenc}
\usepackage{lmodern}
\usepackage{amsmath,amssymb,booktabs,tabularx,array,longtable}
\usepackage{graphicx,xcolor,microtype,listings}
\usepackage[numbers,sort&compress]{natbib}
\usepackage[colorlinks=true,linkcolor=blue,citecolor=blue,urlcolor=blue]{hyperref}
\usepackage{tikz}
\usetikzlibrary{positioning,arrows.meta}
\definecolor{inkblue}{RGB}{35,65,95}
\definecolor{lightblue}{RGB}{241,246,251}
\lstdefinestyle{promptstyle}{basicstyle=\ttfamily\scriptsize,breaklines=true,
  columns=fullflexible,keepspaces=true,frame=single,showstringspaces=false,
  backgroundcolor=\color{lightblue},rulecolor=\color{inkblue},
  literate={’}{{'}}1 {–}{{-}}1 {—}{{---}}3 {“}{{``}}2 {”}{{''}}2}
\newcommand{\system}{Idea2Prior}
\newcommand{\code}[1]{\texttt{\detokenize{#1}}}
\title{Idea2Prior: Auditable Paper Retrieval and Ranking from Research Ideas}
\author{Shitian Yang \quad Ruijie Tao \quad Jian Huang\\
Department of Computer Science, Johns Hopkins University\\
\small \texttt{syang214@jh.edu}, \texttt{rtao12@jh.edu}, \texttt{jhuan243@jh.edu}}
\date{October 13, 2026}

\begin{document}
\maketitle

\begin{abstract}
Finding a paper from an informal research idea requires more than matching topic words: relevant work may use different terminology, and its distinguishing mechanism may be absent from the abstract. We present \system{}, an auditable literature-search system built on PaperArena. It combines multi-source discovery, identity verification, facet-based abstract assessment, semantic citation expansion, and joint Expert ranking. A retrieval-only mode ranks up to five papers from a shortlist of at most 60, using complete available abstracts and locally selected PDF evidence from five fixed acquisition attempts, without generating a long-form report. In a historical comparison covering ten queries from each of EASY, MEDIUM, and HARD, aggregate reported Hit@5 is 53.3\% for \system{}, 43.3\% for ChatGPT web Deep Research, and 36.7\% for Gemini web Deep Research under the documented accounting convention. The advantage is concentrated in EASY and MEDIUM; ChatGPT leads on HARD. Selected \system{} runs average 8 minutes 53 seconds, compared with user-reported web averages of 5.2 and 4.7 minutes. A separate nine-query API sample averages \$0.0945 per task; this is a historical cost observation, not a matched comparison with web subscription costs. Pairwise Top 5 coverage ranges from 6\% to 34\%, showing differing returned evidence sets. The system also exposes optional implementation analysis and citation-evolution tools. These results support targeted idea-to-paper retrieval while leaving general web research, forum search, and broader relevance evaluation open.
\end{abstract}

\section{Introduction}
Researchers often remember an idea before they remember a title: a neighborhood with uncertain boundaries, a circuit transformation with a particular constraint, or an assistant that recognizes when a complex task changes. Recovering relevant prior work from such descriptions is difficult because vocabulary, application setting, and the level of technical detail can differ substantially between a query and a paper. A useful assistant must retrieve candidates, establish their bibliographic identities, and distinguish a close conceptual match from work that merely shares a topic.

LLM-based research systems provide useful ways to explore and synthesize sources. STORM organizes research into multi-perspective information gathering and article outlining~\citep{storm}, Co-STORM supports interactive exploration~\citep{costorm}, and Agent Laboratory connects literature review with experimentation and reporting~\citep{agentlab}. Our focus is narrower: given an idea, which five identifiable papers are most likely to match it, and what evidence explains their selection? This objective is not equivalent to the quality of a general research report.

We introduce \system{}, an extension of PaperArena and its implementation~\citep{paperarena,paperarena-code}. The contribution is an engineering and evaluation framework for inspectable idea-to-paper retrieval: it preserves search queries, candidate identities, assessments, filtering reasons, PDF availability, and the final ranking. This makes it possible to distinguish a discovery miss from a candidate that was retrieved but lost during filtering, or from a task that failed operationally. A high-level result alone cannot make those distinctions.

The central design separates inexpensive assessment of a broad candidate pool from a bounded examination of document evidence. After abstract assessment and citation expansion, an Expert model jointly considers up to 60 candidates. Five papers are selected for PDF acquisition; unavailable PDFs are not replaced by additional papers in retrieval-only mode. Complete available abstracts remain in the Expert input, along with evidence availability and up to six locally selected PDF passages per acquired paper. A missing PDF does not make a paper irrelevant and does not prevent it from ranking first. The optional full-report mode retains downstream synthesis for users who need a written literature analysis.

The revised evaluation covers 30 query cases: ten from each of three prompt tiers, with shared targets across tiers. The reported aggregate Hit@5 is 16/30 for \system{}, 13/30 for ChatGPT web Deep Research, and 11/30 for Gemini web Deep Research. The ChatGPT total includes one convention-based credit, disclosed in Section~\ref{sec:protocol}. \system{} leads both baselines by 20 percentage points on EASY and MEDIUM, but trails ChatGPT by 10 points on HARD. Mean selected-run latency is 8.89 minutes, and a separately reported nine-query cost sample averages \$0.0945. Thus the present evidence shows a recovery advantage in two tiers and a low observed API cost, with longer latency than the supplied browser averages.

Our contributions are:
\begin{enumerate}
\item An auditable retrieval-and-ranking workflow with identity-aware candidate records, semantic citation filtering, and a joint Top 5 decision over a bounded shortlist.
\item A retrieval-only execution mode that preserves abstract and PDF evidence assessment while omitting report-specific processing, exposed through local web, Python, CLI, and Notebook interfaces.
\item A three-system historical evaluation with explicit target-recovery, agreement, latency, token, and cost accounting, together with preserved prompts and outputs.
\item Optional implementation-analysis, citation-evolution, planning, benchmark-landscape, and citation-gap extensions, with their earlier exploratory evaluations distinguished from the retrieval benchmark.
\end{enumerate}

\section{Related Work}
\paragraph{Research agents and scholarly infrastructure.}
STORM, Co-STORM, and Agent Laboratory illustrate complementary approaches to source exploration, synthesis, and research assistance~\citep{storm,costorm,agentlab}. PaperArena supplies a foundation for evaluating tool-augmented reasoning over scientific literature~\citep{paperarena}. \system{} builds on that foundation and studies a specific retrieval endpoint rather than claiming to introduce the general agentic-research setting. Scholarly indexes such as OpenAlex~\citep{openalex} provide useful metadata and links, while paper identity must still be checked across publication and preprint records. Our public retrieval profile uses Semantic Scholar, Crossref, and arXiv-related resolution; OpenAlex remains an optional legacy backend.

\paragraph{Evidence and research artifacts.}
SciFact explicitly associates scientific claims with supporting or refuting evidence~\citep{scifact}. Our task instead ranks works by how well they match an idea; facet scores and selected passages are decision evidence, not formal proof of novelty or claim truth. Implementation analysis similarly surfaces code, pseudocode, dependencies, and missing details without establishing that an entire published experiment is reproducible. These distinctions matter when model-produced assessments are later reused in scientific writing.

\paragraph{Citation-based exploration.}
Citation neighborhoods offer an additional discovery route when a query and its target use different words. A citation edge alone does not establish conceptual relevance or technical inheritance. Connected Papers, for example, describes a similarity graph based on co-citation and bibliographic coupling rather than a direct citation tree~\citep{connectedpapers}. We therefore use external graphs as exploratory references, not as ground truth for causal research lineages.

\section{Method}
\label{sec:method}
\subsection{Task, execution modes, and evidence records}
Given a research idea $q$, the retrieval task returns an ordered set $R_q$ of at most five distinct works. Each entry includes a title, available bibliographic identifiers, and a concise explanation grounded in the supplied evidence. The benchmark allows publications from 2015--2025 and verified preprint versions of the same work. Different versions are consolidated when their identity is established; a predecessor or follow-up is not automatically the same work.

\begin{figure}[t]
\centering
\begin{tikzpicture}[node distance=4mm,box/.style={draw=inkblue,fill=lightblue,
rounded corners,align=center,text width=.87\linewidth,inner sep=5pt,font=\small},
arr/.style={-{Stealth[length=2mm]},draw=inkblue}]
\node[box] (a) {Research idea and constraints\quad $\rightarrow$\quad facets and search queries};
\node[box,below=of a] (b) {Multi-source discovery, identity checks, and abstract assessment};
\node[box,below=of b] (c) {Five citation seeds\quad $\rightarrow$\quad semantic neighbor scoring\quad $\rightarrow$\quad candidate updates};
\node[box,below=of c] (d) {Up to 60 eligible works\quad +\quad five fixed PDF attempts, no replacement};
\node[box,below=of d] (e) {Complete available abstracts + available PDF snippets + assessment/status records};
\node[box,below=of e] (f) {Joint Expert ranking\quad $\rightarrow$\quad up to five works and auditable reasons};
\foreach \x/\y in {a/b,b/c,c/d,d/e,e/f} \draw[arr] (\x)--(\y);
\end{tikzpicture}
\caption{Current retrieval-only workflow. The five citation seeds, five PDF attempts, and final five results are separate selections. The Expert can select any eligible work in the supplied shortlist.}
\label{fig:pipeline}
\end{figure}
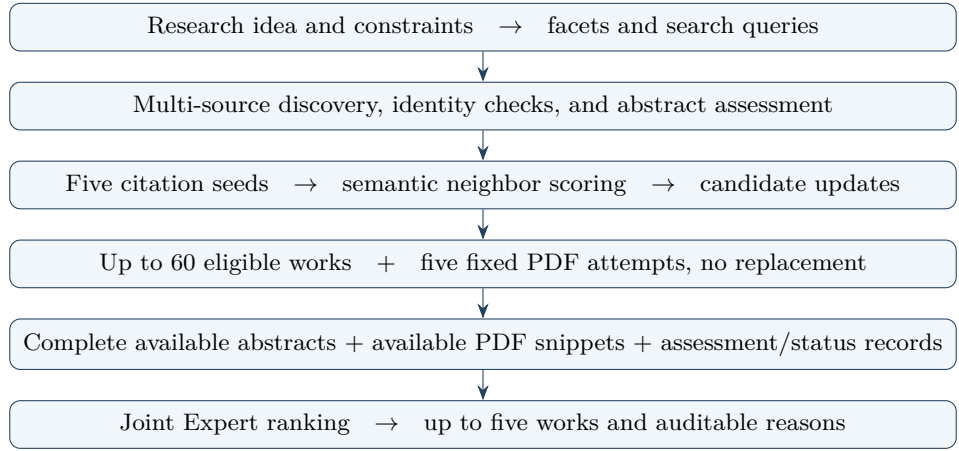

The public system supports \code{retrieval_eval} and \code{full_report}. The former ends after final ranking and audit export. The latter adds report-oriented evidence refinement, method/experiment extraction, representative-paper selection, and synthesis artifacts. Optional research extensions are separately selected tasks; none is automatically included in the Top 5 benchmark. Figure~\ref{fig:pipeline} summarizes the retrieval-only path.

\subsection{Idea structuring, discovery, and identity verification}
The pipeline preserves the original request and derives 3--7 technical facets, structural cues such as datasets and method families, and higher-level problem/setting/mechanism/assumption facets. These generated interpretations aid search and assessment; they do not replace the original idea. Model-recalled seeds and web-derived paper candidates are discovery hypotheses, requiring bibliographic resolution before they can be treated as established identities.

The current profile combines recalled seeds, web search, and scholarly metadata. Semantic Scholar supplies paper records and citation neighborhoods. Crossref supplies bibliographic verification and metadata; merely receiving a reference in a Crossref response does not insert it into the candidate pool. Explicit Crossref-reference expansion is disabled in the default profile. arXiv identifier resolution and PDF access remain available even when arXiv keyword search is disabled. OpenAlex is disabled in this profile but retained for compatible older workflows. Consequently, provider availability and a paper's route into the candidate registry are recorded separately.

Local selection bounds the candidate workload. The protected relevance policy uses a BM25-like score over title and complete available abstract text, complemented by recorded retriever ranks. It reserves limited capacity for corroborated seeds, strong retrieval ranks, and missing-abstract candidates, and handles publication-year eligibility and duplicate entities explicitly. This is still a budgeted selection algorithm: a relevant candidate can be dropped below a capacity boundary. It is not a guarantee of recall, and it is distinct from the semantic citation-neighbor gate.

\subsection{Abstract assessment and semantic citation expansion}
The merged abstract prompt produces an ordinal overlap label, covered facet IDs, task and method similarities, a short reason, and the paper's high-level facets. The labels are $3$ for high overlap, $2$ for partial overlap, $1$ for adjacent work, and $0$ for no overlap. Missing evidence and invalid model outputs retain explicit unassessed or failed states; they are not converted into valid zero scores. In the staged profile, the older lexical-zero and Boolean LLM prefilters are bypassed before abstract scoring. Local budget selection still applies.

After initial assessment and metadata enrichment, up to five papers seed citation expansion through Semantic Scholar. Unique neighbors with abstracts are assessed in batches of up to 20. The Light-role model receives the original research query, complete idea, generated search queries, organized keywords, facets, and each neighbor's title and abstract. Its rubric recognizes synonyms and mechanism-level matches, and explicitly forbids assigning zero merely because a keyword is absent. The complete prompt and hypothetical calibration examples appear in Appendix~\ref{app:prompts}.

Valid zero-score neighbors are excluded; positive neighbors are ordered $3>2>1$, with up to 15 retained per seed under the default setting. Missing abstracts or scoring failures remain unassessed and bypass that positive-only cap, while still being subject to the global candidate budget. The default global budget admits at most 50 new citation candidates. Identity checks, deduplication, and abstract assessment then update the pool. Neither semantic filtering nor a citation relationship grants automatic access to the final shortlist.

\subsection{Shortlisting and bounded PDF evidence}
\label{sec:pdf}
The current profile selects up to 50 assessed works plus a buffer of up to ten boundary or uncertain candidates. Thus at most 60 distinct eligible works reach the Expert, and the actual count can be smaller. Assessment ordering considers overlap grade, task and method similarity, covered facets, and related structural signals; the staged policy does not use citation count as a relevance tiebreaker. Conflicting identities and known out-of-range publication years are excluded and logged.

Exactly five candidate slots, or fewer when fewer eligible candidates exist, are selected for PDF acquisition. These are fixed paper-level attempts, not a promise of five successful PDFs or a cap of five HTTP requests. If a PDF is unavailable, the workflow does not substitute another paper. The attempted paper remains available through its abstract or metadata, and the acquisition outcome is included in the Expert input.

For each available PDF, the default local extractor scans native text from all pages using windows of at most 1,800 characters, with a half-window stride. It orders windows by distinct idea/facet terms, capped term frequency, section cues, and deterministic location tiebreakers. It first takes a high-scoring window, then seeks nonoverlapping query-relevant coverage of method, experiment, and conclusion cues before filling the remaining slots. At most six windows are sent per paper, in reading order. This selector is a lexical window heuristic with section cues: it is not a guaranteed Introduction/Related Work/Method/Conclusion parser, and it makes no extra model call. Full extracted text stays in local artifacts. Sent windows record page locations, offsets, and hashes, and are explicitly marked partial.

\subsection{Joint Expert ranking and failure handling}
The Expert receives the entire eligible shortlist, complete available abstracts, existing assessments, and only the PDF material actually acquired and selected. Earlier grades are evidence rather than a mandatory final ordering. Papers with unavailable or unattempted PDFs may rank first. The model returns exactly $\min(5,|C_q|)$ candidate IDs in consecutive rank order, a one- or two-sentence explanation for each, and an evidence-source label. Reasons are qualitative; the system does not claim calibrated probabilities.

Validation checks candidate membership, duplicate identities, rank order, and consistency between claimed evidence sources and supplied material. A failed Expert output is recorded as a ranking failure rather than replaced silently with a rule-based Top 5. Operational completion and scientific target recovery are separate outcomes. A partial run may contain a valid ranking while recording unavailable documents or incomplete checks; a successful ranking alone does not prove a target hit.

\subsection{Efficiency and optional report generation}
Retrieval-only optimizations include concurrent independent seed branches, reuse of valid unchanged model responses within a run, batched scholarly requests, shared HTTP connections, compact Expert payloads, and skipping report-only evidence processing. Reuse keys include the prompt, evidence, model, and configuration dependencies; changed evidence must be reassessed. Compact serialization preserves the abstract and selected evidence while removing duplicated or audit-only material. These choices reduce redundant work by design, but this paper does not report a controlled ablation establishing a latency or accuracy effect for each optimization.

The Expert has a separate maximum output budget of 16,384 tokens. Its prompt still requests only up to five concise entries; this ceiling is not observed consumption or a request for a long reasoning transcript. Request budgets, deadlines, status records, cancellation, and provider-specific configuration are available through all entry points. The public runner does not automatically restart a whole task. In full-report mode, additional artifacts include method/experiment snippets, representative-paper selections, synthesis input, and a rendered report; these costs are outside the reported retrieval-only comparison.

\section{Optional Research Extensions}
\label{sec:extensions}
\subsection{Code Extractor}
Code Extractor takes a supplied PDF, Markdown, or text document and emits structured implementation diagnostics and a Markdown report. It combines rule-based code/pseudocode extraction with optional LLM analysis of completeness, frameworks, dependencies, model structure, and missing components. Native PDF text extraction is the default; structured MinerU parsing is explicitly optional. Model-recovered descriptions must be distinguished from verbatim executable source.

The public interface separates analysis from execution. With model assistance disabled, scores are labeled heuristic; with code execution disabled, execution outcomes remain unrun even if syntax checks succeed. Missing packages, parser failures, and requested model-analysis failures are exposed as diagnostics. This tool can identify potentially missing details, but successful parsing, syntax checks, or high model scores do not establish paper-level reproducibility.

\subsection{Evolution Mapper and Planner--Worker}
Evolution Mapper retrieves a topic's papers and citation neighbors, builds a directed graph, groups papers by year, and exports Mermaid diagrams, key-paper summaries, and path analyses. Citation edges and chronological views are descriptive evidence. They must not be interpreted as proof of technical influence, and edge orientation must be distinguished from chronological traversal. The paper limit bounds seed discovery; citation neighbors can enlarge the graph.

Planner--Worker coordinates restored analysis tools with argument validation, dependency-aware execution, step outcomes, and optional Critic review. Public runs bound the number of steps and set automatic re-execution to zero. A Critic request to repeat work is preserved rather than executed automatically. A supplied document is bound to a fixed tool input to avoid ambiguous file selection.

\subsection{Benchmark landscape, citation gaps, and visual comparison}
The optional benchmark-landscape module summarizes metadata-derived datasets, metrics, and method families. Its output is a preliminary exploration rather than a verified systematic survey. Citation-gap analysis evaluates potential references against a draft excerpt and recognized existing references. These extensions retain their earlier discovery workflow and are not part of the Top 5 retrieval evaluation.

The earlier manuscript also described cross-paper figure/table extraction, multimodal similarity retrieval, and pairwise visual comparison. We retain that design and its reported operational pilot in Appendix~\ref{app:legacy}, but its implementation and run files have not been located in the audited project archive. It is therefore not presented as a currently callable or reproduced component of the public release.

\section{Experiments}
\subsection{Dataset and historical comparison protocol}
\label{sec:protocol}
The released benchmark contains 30 queries in each of EASY, MEDIUM, and HARD. According to the dataset construction record, GPT-6 Astra in ChatGPT was used to extract and generate the shared literature-search prompts. EASY supplies more identifying detail, MEDIUM removes some detail, and HARD uses shorter, less specific descriptions. These are intended difficulty levels rather than independently calibrated labels. The tiers are not independent paper pools; in particular, MEDIUM and HARD share targets with different wording.

The present comparison uses the first ten IDs of each tier: EASY N001--N010, MEDIUM C001--C010, and HARD C001--C010. Each system receives the same filled item prompt: a paper-finding instruction, the 2015--2025 publication/verified-version constraint, up-to-five distinct works with exact bibliographic fields and a short match explanation, identity verification and deduplication requirements, and the idea text. The designated target is not supplied to retrieval. Original prompts and outputs are preserved; Appendix~\ref{app:eval-prompts} gives the exact shared wrappers and a three-tier example. These are retrieval prompts, distinct from the internal scoring prompts.

ChatGPT and Gemini were tested through their October 2026 web Deep Research interfaces; backend model/build identifiers and subscription tiers were not recorded. \system{} used its API pipeline. Preserved Expert outputs identify the configured alias \code{openai/gemini-3.8-flash}; a compatibility-protocol alias does not independently verify the upstream served model. The released native-Gemini preset is not a newly benchmarked replacement for those historical runs.

Results use final selected runs, including authorized replacements of operationally failed attempts. They therefore measure selected-result recovery, not first-attempt reliability. EASY initially produced five target hits over ten planned queries before recovery; the final selection produced nine. We retain the historical ChatGPT EASY convention of 7/10: six observed hits plus one convention-based counted hit. No returned paper or intersection is fabricated for that credit. Counting only observed hits would make ChatGPT EASY 6/10 and aggregate recovery 12/30; our primary table preserves the earlier agreed convention explicitly.

These browser outputs also carry a control limitation: new conversations were used, but the operator reported ChatGPT memory/history references enabled during earlier testing. The adopted HARD comparison uses the replacement NEW report. This historical record does not establish a newly controlled, contamination-free evaluation. Future comparisons should use isolated sessions with memory disabled and a frozen protocol.

\subsection{Metrics}
For $n$ planned queries with designated targets $g_q$, target recovery is
\begin{equation}
\operatorname{Hit@5}=\frac{1}{n}\sum_{q=1}^{n}\mathbf{1}[g_q\in R_q].
\end{equation}
Identity matching uses normalized persistent identifiers when available and reviewed title/version associations otherwise. Hit@5 measures recovery of a known positive; it does not assert that other returned works are irrelevant. Strict designated-target recovery should not be silently replaced by a broader work-family match.

To examine how results depend on topics and prompt wording, we also measure pairwise agreement. For two systems $A,B$ in a ten-query tier,
\begin{equation}
\operatorname{Coverage}(A,B)=\frac{\sum_{q=1}^{10}|R_q^A\cap R_q^B|}{10\times5}.
\end{equation}
Only known same-query paper identities count, ignoring rank. The denominator remains 50 planned slots even when fewer papers are returned; an unfilled slot contributes no observed match. This is symmetric fixed-slot coverage, not Jaccard similarity or a denominator based on a variable returned-list size. Agreement complements target recovery but cannot establish that the benchmark is unbiased or representative.

\subsection{Target recovery and cross-system coverage}
\begin{table}[t]
\centering\small
\begin{tabular}{lrrr}
\toprule
Tier & \system{} & ChatGPT web DR & Gemini web DR\\
\midrule
EASY & \textbf{90\% (9/10)} & 70\% (7/10)$^{\dagger}$ & 70\% (7/10)\\
MEDIUM & \textbf{50\% (5/10)} & 30\% (3/10) & 30\% (3/10)\\
HARD & 20\% (2/10) & \textbf{30\% (3/10)} & 10\% (1/10)\\
\midrule
All query cases & \textbf{53.3\% (16/30)} & 43.3\% (13/30)$^{\dagger}$ & 36.7\% (11/30)\\
\bottomrule
\end{tabular}
\caption{Historical target Hit@5. $^{\dagger}$ChatGPT EASY includes one convention-based credit; Section~\ref{sec:protocol} discloses the observed-only alternative. All rows use ten planned queries per tier. Aggregation is over query cases, not 30 independent target papers.}
\label{tab:hit}
\end{table}

Table~\ref{tab:hit} shows a higher aggregate reported recovery rate for \system{}, with a 10.0-point difference from ChatGPT and a 16.7-point difference from Gemini. This advantage is not uniform: ChatGPT ranks above \system{} on HARD. The small, overlapping target set and historical protocol do not support a broad or statistically established superiority claim.

\begin{table}[t]
\centering\small
\begin{tabular}{lrrr}
\toprule
Pair & EASY & MEDIUM & HARD\\
\midrule
\system{}--ChatGPT & 28\% (14/50) & 22\% (11/50) & 6\% (3/50)\\
\system{}--Gemini & 30\% (15/50) & 12\% (6/50) & 16\% (8/50)\\
ChatGPT--Gemini & 34\% (17/50) & 20\% (10/50) & 14\% (7/50)\\
\bottomrule
\end{tabular}
\caption{Mutual Top 5 coverage with a fixed denominator of 50 slots per tier. The numerator counts actual known intersections and is independent of the convention-based target credit.}
\label{tab:coverage}
\end{table}

Coverage ranges from 6\% to 34\% (Table~\ref{tab:coverage}). Thus the systems often return different papers for the same idea. This motivates inspecting complementary evidence sets, but low agreement alone does not show that their union is more relevant or that a particular system searches more broadly.

\subsection{Runtime, reported tokens, and material availability}
\begin{table}[t]
\centering\small
\setlength{\tabcolsep}{5pt}
\begin{tabular}{lrrrr}
\toprule
Scope & Mean time & Mean input & Mean output & Mean total\\
\midrule
EASY (10) & 8m54s & 288,821.8 & 88,153.6 & 376,975.4\\
MEDIUM (10) & 9m49s & 304,122.9 & 91,695.7 & 395,818.6\\
HARD (10) & 7m57s & 302,838.2 & 88,621.0 & 391,459.2\\
All (30) & \textbf{8m53s} & 298,594.3 & 89,490.1 & 388,084.4\\
\bottomrule
\end{tabular}
\caption{Selected \system{} runs. Token counts are provider-reported subtotals summed across model stages, not complete consumption.}
\label{tab:efficiency}
\end{table}

The all-tier mean is 533.413 seconds (Table~\ref{tab:efficiency}). ChatGPT and Gemini web means are operator-reported at 5.2 and 4.7 minutes per query, respectively, without independent per-query logs. \system{} is slower under these reported averages. Our timing includes failed requests within each selected run but excludes superseded whole-query attempts and shared preflight checks. An all-attempt EASY accounting gives 13.04 minutes per planned query, illustrating the effect of operational recovery on total effort.

Reported model usage totals 11,642,532 tokens across the 30 selected runs, averaging 388,084.4 tokens per query. However, 1,491 requests lack usage records: 754 in EASY, 63 in MEDIUM, and 674 in HARD. Missing usage is unknown, not zero, and does not itself prove a charge. These subtotals exclude Codex editing usage and do not reconstruct a complete bill. Browser baseline token consumption is unavailable.

All 30 selected runs contain successful Expert rankings while their overall workflow status is \code{PARTIAL}. The preserved material summaries record 31/50 available PDFs for EASY, 31/50 for MEDIUM, and 29/50 for HARD: 91/150 fixed acquisition attempts overall. This supports the need to rank with heterogeneous evidence availability. It does not establish the causal effect of full-text evidence, and the PDF counts are not a complete explanation of every partial-status diagnostic.

\subsection{Cost per task and official API references}
\label{sec:cost}
A separate operator-reported nine-query API sample averages \textbf{\$0.0945 per task}. Its cohort differs from the 30-query timing/token cohort, and the observation is not a guaranteed public tariff or a measured browser-baseline cost. Costs depend on the endpoint, selected models, tools, and unreported usage.

For context, Table~\ref{tab:cost} reports official Deep Research API references checked on October 13, 2026. Repricing our known uncached token workload uses
\begin{equation}
\widehat C(r_{\mathrm{in}},r_{\mathrm{out}})=
\frac{298{,}594.3\,r_{\mathrm{in}}+89{,}490.1\,r_{\mathrm{out}}}{10^6}.
\end{equation}
The OpenAI figures exclude tools and unreported tokens and sum many requests, not one model context. They do not predict another agent's actual token use. Official documentation lists July 23, 2026 as the shutdown date for the two named OpenAI Deep Research models~\citep{openai-dr}; their published rates are retained here as historical references, not currently available API recommendations.

\begin{table}[t]
\centering\small
\begin{tabularx}{\linewidth}{@{}p{.27\linewidth}Xr@{}}
\toprule
Cost reference & Basis & USD/task\\
\midrule
\system{} & Separate historical nine-query sample & 0.0945\\
OpenAI o3 Deep Research & Historical \$10 input / \$40 output per million; known-workload illustration & 6.57\\
OpenAI o4-mini Deep Research & Historical \$2 input / \$8 output per million; known-workload illustration & 1.31\\
Gemini Deep Research & Google's typical-task preview estimate & 1--3\\
Gemini Deep Research Max & Google's typical-task preview estimate & 3--7\\
\bottomrule
\end{tabularx}
\caption{Different cost accounting bases, not measured costs of matched baseline runs. OpenAI references use published historical model rates~\citep{o3-price,o4-price}; Google ranges describe its example tasks~\citep{gemini-dr}.}
\label{tab:cost}
\end{table}

Google's ranges are variable preview estimates, not fixed per-report tariffs and not estimates derived from our aggregate tokens. A comparable Gemini agent calculation would require model-specific usage, caching, and tool counts. Our observed sample mean is numerically below these references, but no matched cost-effectiveness ranking follows. The accuracy evaluation used ChatGPT and Gemini \emph{web} Deep Research; their API token usage and backend models were not measured, and subscription prices cannot be treated as observed per-query bills.

\subsection{Code Extractor evaluation}
We retain the separate ten-paper implementation-analysis experiment from the manuscript. Its reference annotations cover pseudocode completeness (0--2), reproducibility level (1--3), frameworks, dependencies, model structure, and missing implementation components. An LLM judge compares each tool output with the annotations on 0--10 agreement scales. These are judge-assessed comparison scores rather than execution-based reproducibility measurements. Detailed per-paper results are provided in Appendix~\ref{app:legacy}.

The mean overall comparison score is 6.9/10. Pseudocode agreement averages 7.3/10, reproducibility-level agreement 5.0/10, and model-structure agreement 6.0/10. The per-paper missing-component scores average 8.65/10, with five papers receiving 10/10. Dependency agreement is 10/10, but all ten reference dependency lists are empty, so this result does not demonstrate recovery of nonempty dependencies or correct framework identification. For example, the Transformer case receives 10/10 for missing-component agreement but only 5/10 for reproducibility-level agreement. The evaluation supports diagnostic utility in some dimensions while exposing weaker calibration and a limited dependency test set.

\subsection{Evolution Mapper evaluation}
The separate evolution-map study covers Transformers, generative adversarial networks, and ImageNet/CNN architectures. Each topic uses a Connected Papers reference set of 41 publications and a canonical seed paper. A model judge assesses coverage, citation relationships, network similarity, and additional value on 0--10 scales. Connected Papers is an exploratory similarity reference, not verified ground truth for direct citation edges or technical inheritance~\citep{connectedpapers}. Accordingly, the scores below are comparative judgments rather than measured citation precision.

\begin{table}[t]
\centering\small
\begin{tabular}{lrrrrr}
\toprule
Topic & Overall & Coverage & Citation & Network & Add. value\\
\midrule
Transformers & 6.5 & 5.5 & 6.5 & 5.5 & 8.5\\
GAN & 6.8 & 6.0 & 6.5 & 6.0 & 8.5\\
ImageNet/CNN & 7.5 & 7.0 & 8.0 & 7.0 & 9.0\\
\midrule
Mean & 6.9 & 6.2 & 7.0 & 6.2 & 8.7\\
\bottomrule
\end{tabular}
\caption{The manuscript's separate Evolution Mapper experiment; all scores are model-judge ratings out of ten. Displayed means are rounded to one decimal.}
\label{tab:evolution}
\end{table}

The reported overall mean is 6.9/10, with ImageNet/CNN obtaining the highest topic score. Additional-value ratings average 8.7/10, while coverage and network similarity average 6.2/10. The broader graphs can help exploration yet omit influential papers or contain weakly relevant neighbors. These findings motivate better discovery and relation validation; they do not establish universal superiority over the reference tool. No run-level standard deviations are available for the table, so none are claimed.

\subsection{Cross-paper visual comparison pilot}
The manuscript's visual-comparison pilot used nine PDFs, six query intents, and visual, semantic, and combined retrieval modes (18 comparisons). Eight PDFs were processed successfully, producing 137 figures, of which 134 had captions (97.8\%). This is 17.1 figures per successfully processed PDF, or 15.2 per attempted PDF. Twelve comparisons completed and six failed, mainly in association with unavailable content lists; mean latency was 6.73 seconds per successful comparison, with five retrieved figures on average.

These are parsing, availability, and latency measurements, not retrieval precision or semantic-comparison accuracy. The design and pilot are retained in Appendix~\ref{app:visual}. The current public release does not expose this separate implementation, and the pilot was not rerun during the present retrieval evaluation. Hand-labeled figure relevance and validated numeric comparisons remain necessary to evaluate its scientific utility.

\subsection{Public-interface validation}
The public-interface restoration was checked with offline regression tests, an actual browser document upload, an actual Notebook kernel, and an installed-wheel local trial. These checks validate plumbing, diagnostic reporting, and local execution paths. They do not constitute a fresh scientific benchmark of the extensions. Paid model trials for the restored extensions, the full MinerU path, and a new full-report end-to-end experiment were not run as part of this revision.

\section{Discussion and Limitations}
\paragraph{Recovery, evidence, and interpretability.}
The results favor \system{} on EASY and MEDIUM while showing substantial residual difficulty on HARD. Explicit intermediate records make failures inspectable: discovery, identity resolution, budget selection, semantic filtering, PDF acquisition, and final ranking are distinct potential causes. Historical filtering failures motivated the semantic neighbor gate, but the present comparison does not isolate that change's contribution through an ablation. It would be incorrect to conclude that all remaining misses come from filtering rather than discovery.

\paragraph{Evaluation scale and controls.}
The 30 query cases include overlapping targets and a manually assembled distribution; the first ten items in each tier are not a random sample of all scientific literature. Benchmark authoring used an LLM, browser build identifiers were unavailable, history contamination cannot be excluded retrospectively, and one baseline credit follows an explicit historical convention. These factors limit causal and statistical interpretation. Future evaluation should freeze inputs, use isolated sessions, report first-attempt and recovered performance separately, and obtain independent relevance judgments for every returned work.

\paragraph{Efficiency and cost.}
The retrieval-only design omits report generation and bounds full-text acquisition, but the reported API workflow still takes longer than the operator's web averages. Unknown usage and the separate cost sample prevent full token/billing reconstruction. Controlled same-query, same-provider experiments are needed before claiming that a particular optimization improves speed without reducing accuracy, or that one system offers a measured cost-quality advantage over another.

\paragraph{Evidence incompleteness and optional tools.}
Abstracts and a few passages can omit the distinguishing contribution of a paper. Native PDF text extraction does not capture every image or table. Generated overlap labels are not calibrated relevance probabilities, and a model's reproducibility assessment is not an execution-based replication study. Historical extension evaluations include empty dependency labels, model-judge variation, and an external similarity graph used as a comparison reference. Visual comparison remains an unreproduced historical component.

\paragraph{Scope.}
Our comparison addresses finding papers from research ideas. It does not evaluate general website exploration, forum search, comprehensive web research, report quality, legal patent novelty, or the correctness of every citation recommendation. General web and forum capabilities remain limited. A designated-target miss can coexist with useful related papers, so target Hit@5 should be complemented by independently adjudicated relevance and usefulness.

\section{Conclusion}
\system{} extends PaperArena with an auditable workflow for retrieving and ranking papers from research ideas. Joint ranking combines an abstract-level shortlist with bounded, explicitly partial PDF evidence, while a retrieval-only mode avoids report-specific work. In the documented historical comparison, aggregate reported Hit@5 is 53.3\%, exceeding the two browser baselines overall, with a clear HARD-tier exception and longer reported latency. A separate historical cost sample averages \$0.0945 per query. These findings motivate focused scholarly retrieval with inspectable evidence; they do not establish general Deep Research superiority, calibrated novelty assessment, or a matched baseline cost advantage.

\clearpage
\appendix

\section{Shared Evaluation Inputs}
\label{app:eval-prompts}
The following wrappers are reproduced from the packaged EASY N001 and MEDIUM C001 prompt files. HARD uses the same wrapper as MEDIUM. Only the idea text changes across items. These are the original user-facing retrieval inputs, not instructions to reveal a designated answer. The original files and hashes are retained under \code{datasets/<tier>/prompts/}; gold metadata and prior results are evaluated separately and are not runtime inputs.
\subsection{EASY wrapper}

\begin{lstlisting}
Search for the research paper that best matches the idea below. The target has a journal, conference, or workshop publication dated from 2015 through 2025, inclusive. A verified preprint version of the same work is acceptable; its date and title may differ from the published version.

Return a ranked list of up to five distinct candidate papers. For each, give the exact title, authors, publication year, venue, a DOI or stable paper URL, and one sentence explaining the match. Rank by the complete idea rather than shared topic words. Verify each paper's identity against an authoritative record. Do not invent citations. Versions of the same work count as one candidate, but a predecessor or follow-up is not automatically the same paper. If no plausible candidate can be verified, say so.

Research idea:
{idea_text}
\end{lstlisting}

\subsection{MEDIUM and HARD wrapper}

\begin{lstlisting}
Search for the research paper that best matches the idea below. The target has a journal, conference or workshop publication dated from 2015 through 2025, inclusive. A verified preprint of the same work is acceptable; its title or date may differ from the published version.

Return up to five distinct candidate works, ranked by the complete idea rather than shared topic words. For each candidate, give its exact title, authors, publication year, venue, DOI or stable paper URL, and one sentence explaining the conceptual match. When a conference paper and an expanded journal article are part of the same work family, identify their relationship and list the versions within one candidate rather than filling multiple ranks with them. Do not merge distinct follow-ups or merely related papers.

Verify identities against authoritative records. Do not invent citations. If several papers fit, state the ambiguity rather than inventing a unique answer. If no plausible candidate can be verified, say so.

Research idea:
{idea_text}
\end{lstlisting}

\subsection{One target expressed at three information levels}

\paragraph{EASY.}

\begin{lstlisting}
Could we infer a neighborhood’s fuzzy edges from residents’ contributed locations rather than force it into a single official boundary? I want maps to distinguish its core from places that only partly belong.
\end{lstlisting}

\paragraph{MEDIUM.}

\begin{lstlisting}
I keep wondering whether a neighborhood has to have a clear edge on a map. Could we represent the way people loosely agree about a place without forcing one exact boundary?
\end{lstlisting}

\paragraph{HARD.}

\begin{lstlisting}
Maybe maps should leave room for disagreement about where a neighborhood begins and ends.
\end{lstlisting}

This example illustrates reduced identifying information; it is not a calibration proof for every query. Broad descriptions can legitimately fit several works. The designated positive is used for post-retrieval Hit@5, while relevance of the other returned works requires separate assessment.

\section{Current Internal Prompt Library}
\label{app:prompts}
This appendix records the current implementation snapshot, checked on October 13, 2026. Templates below are extracted from source string literals; braces denote runtime fields or Python expressions. They document the released workflow, and do not assert that every historical run used an identical implementation snapshot. The preserved user-facing benchmark prompts above are unchanged. Role names describe routing, not a universal model capability tier. Local selection and snippet extraction are algorithms, not additional model prompts.

The old manuscript's Boolean relevance gate and lexical-zero citation filter are not described as active defaults: the staged retrieval profile bypasses the former, and the semantic prompt below replaces the latter. Report-only prompts are explicitly separated from the retrieval-only path. Source locations and SHA-256 fingerprints are recorded in \code{paper/revision_manifest.json}.

\subsection{P01: Idea-to-facets decomposition}

\noindent Role: Light. Source function: \code{decompose_idea_into_facets}.

\begin{lstlisting}
You are a JSON generator. **Do NOT call any tools or write/run any code.**
Think silently and then output the final answer as JSON only.
Return raw JSON only, without backticks or any surrounding text.

Given the following research idea, please extract 3-7 key technical facets that characterize this idea.

A "facet" is a distinct technical component, method, assumption, or contribution of the idea.

Research Idea:
{idea_description}

Please output ONLY a JSON list of strings, where each string is one facet. Example format:
["facet 1", "facet 2", "facet 3"]

JSON output:
\end{lstlisting}

\subsection{P02: Structural feature extraction}

\noindent Role: Light. Source function: \code{extract_idea_features}.

\begin{lstlisting}
You are a JSON generator. **Do NOT call any tools or write/run any code.**
Think silently and then output the final answer as JSON only.
Return raw JSON only, without backticks or any surrounding text.

Given the following research idea, please extract the key structural features:
1. Datasets (e.g., KITTI, nuScenes, ImageNet)
2. Metrics (e.g., mAP, F1, Accuracy)
3. Method Families (e.g., Transformer, CNN, Diffusion)

Research Idea:
{idea_description}

Please output ONLY a JSON object with keys "datasets", "metrics", "method_families". Each value should be a list of strings.
Example: {"datasets": ["KITTI"], "metrics": ["mAP"], "method_families": ["Transformer"]}

JSON output:
\end{lstlisting}

\subsection{P03: High-level idea facets}

\noindent Role: Light. Source function: \code{extract_high_level_facets_from_idea}.

\begin{lstlisting}
You are a JSON generator. **Do NOT call any tools or write/run any code.**
Think silently and then output the final answer as JSON only.
Return raw JSON only, without backticks or any surrounding text.
Do NOT include any comments, trailing commas, or extra tokens. Only output valid JSON.

Given the following research idea, please extract high-level structural facets.
    
Facet Types:
1. problem_facets: What specific problem/task is it solving? (e.g., "monocular 3D detection", "long-horizon sequence modeling")
2. setting_facets: What is the setting, modality, or protocol? (e.g., "indoor RGB-D scenes", "multi-camera surround view")
3. mechanism_facets: What are the key mechanisms or frameworks? (e.g., "camera-space rotation augmentation", "SSM-based state space modeling")
4. assumption_facets: What are the key assumptions? (e.g., "calibrated pinhole camera", "no depth supervision")

Research Idea:
{idea_text}

Please output ONLY a JSON object with keys "problem_facets", "setting_facets", "mechanism_facets", "assumption_facets".
Each value should be a list of short phrases (3-7 phrases max per category).

JSON output:
\end{lstlisting}

\subsection{P04: Model-recalled paper seeds}

\noindent Role: Expert. Source function: \code{seed_papers_from_llm}.

\begin{lstlisting}
You are an expert researcher.

Your task is to recall real academic papers that are closely related
to the following research idea:

"""{idea_description}"""

Requirements:
- Only include real papers that you reasonably believe exist.
- Prefer papers that describe methods, models, or techniques that
  would be informative when evaluating the novelty of this idea.
- If you are uncertain about a detail (e.g., exact year or venue),
  you may set that field to null.

For each paper, provide:
- title: the paper's title (string)
- venue: conference or journal name, or null if unknown
- year: integer publication year, or null if unknown
- doi: string or null
- arxiv_id: string (e.g., "2104.12345") or null
- url: string or null (e.g., publisher or arXiv URL)

Return ONLY raw JSON, no backticks, in the following format:
[
  {
    "title": "...",
    "venue": "...",
    "year": 2024,
    "doi": "10.xxx/yyy" or null,
    "arxiv_id": "2312.00752" or null,
    "url": "https://..." or null
  },
  ...
]

Constraints:
- At most {max_seeds} papers.
- Do NOT include any commentary outside the JSON.

JSON:
\end{lstlisting}

\subsection{P05: Web search query generation}

\noindent Role: Expert. Source function: \code{generate_serpapi_queries}.

\begin{lstlisting}
You are an expert researcher.

Your task is to design web search queries that will help find
academic papers related to the following research idea:

"""{idea_description}"""

Requirements:
- Each query MUST be suitable for Google or DuckDuckGo web search.
- Each query should look like what a researcher would actually type,
  combining key technical phrases from the idea.
- Focus on techniques, methods, or key concepts mentioned in the idea.
- Do NOT restrict the domain explicitly (e.g., do not add generic field names
  unless they clearly appear in the idea itself).
- Do NOT add site filters such as "site:arxiv.org" or date filters.
  These will be handled later by the system.
- **IMPORTANT: Do NOT use quotation marks ("") in queries. Write plain keywords
  without exact-match operators, as they severely limit search results.**
- Avoid overly generic queries (e.g., "deep learning method").
- Aim for diverse queries that cover different angles of the idea.
- At most {max_queries} queries.

Respond with ONLY valid JSON, no backticks:
{"queries": ["query1", "query2", ...]}
\end{lstlisting}

\subsection{P06: SerpAPI result parsing (optional provider)}

\noindent Role: Light. Source function: \code{parse_serpapi_results_with_llm}.

\begin{lstlisting}
You are a helpful research assistant.

You are given:
- A research idea: {idea_description}
- A web search query: {query}

You also have search results (from multiple modes), in JSON:
```json
{results_str}
```

Your task:
- From these results, identify up to {max_papers} entries that are MOST LIKELY to be academic papers relevant to the idea.
- Prefer results that clearly look like academic papers (e.g., conference proceedings, journal articles, arXiv papers, open-access repositories).
- For each selected result, output:
  - title: string (the paper title, cleaned up)
  - year: integer if you can guess from the snippet/title/URL, else null
  - venue: short string (e.g., conference name, journal name, "arXiv"), or null if unknown
  - url: the link
  - source_mode: which search mode it came from (one of: all_non_arxiv, all_arxiv, range_non_arxiv, range_arxiv)
  - reason: 1-2 sentences why this seems relevant to the idea.

Output format:
Return ONLY a JSON array of objects with the above fields. No extra commentary.

JSON:
\end{lstlisting}

\subsection{P07: Web result parsing}

\noindent Role: Light. Source function: \code{parse_web_results_with_llm}.

\begin{lstlisting}
You are a helpful research assistant.

You are given:
- A research idea: {idea_description}

You also have web search results in JSON:
```json
{results_str}
```

Your task:
- From these results, identify up to {max_papers} entries that are MOST LIKELY to be academic papers relevant to the idea.
- Prefer results that clearly look like academic papers (conference proceedings, journal articles, arXiv papers).
- For each selected result, output:
  - title: string (the paper title, cleaned up)
  - year: integer if you can guess from the snippet/title/URL, else null
  - venue: short string (e.g., conference name, journal name, "arXiv"), or null if unknown
  - url: the link
  - source_mode: the source_mode from the input (e.g., serpapi_all_years, ddg_arxiv)
  - reason: 1-2 sentences why this seems relevant to the idea.

Output format:
Return ONLY a JSON array of objects with the above fields. No extra commentary.

JSON:
\end{lstlisting}

\subsection{P08: Bibliographic query generation}

\noindent Role: Expert. Source function: \code{generate_search_queries_from_idea}.

\begin{lstlisting}
You are an expert researcher.

Your task is to design bibliographic search queries for academic
paper databases (e.g., OpenAlex, Semantic Scholar) that are relevant
to the following research idea:

"""{idea_description}"""

Requirements:
- Each query should be a short keyword-style query (not a full sentence).
- Combine key technical terms from the idea, including:
  - problem/task,
  - key methods or mechanisms,
  - important properties or techniques.
- Do NOT add year or venue constraints in the query.
  (The system will handle year filters separately.)
- Do NOT add site filters (e.g., "site:arxiv.org").
- Queries should be reasonably distinct to cover different angles.

CRITICAL RULES for query formatting:
- Each query should use at MOST ONE exact phrase in quotes
- Remaining keywords should be unquoted, separated by spaces
- Avoid multiple quoted phrases in the same query (this kills recall)
- Mix broad queries (core concept) with more specific queries (concept + 1-2 keywords)
- Keep queries simple and focused

Return ONLY raw JSON list of strings:
["query 1", "query 2", ...]

Constraints:
- At most {num_queries} queries.
- Do NOT include any commentary outside the JSON.

JSON:
\end{lstlisting}

\subsection{P09: Diversified seed queries}

\noindent Role: Expert. Source function: \code{generate_seed_queries_llm}.

Year hints in this captured template are inherited query-generation examples; publication eligibility is enforced separately by the configured year range.

\begin{lstlisting}
You are generating search queries for academic paper retrieval.

Research topic: {topic}
Research idea: {idea[:600]}

Key facets:
{facet_text}

TASK: Generate {max_queries} short, diversified search queries suitable for Google Scholar / arXiv / Semantic Scholar.

Guidelines:
- Keep each query SHORT (3-8 words, like search engine queries)
- Cover different angles: task/problem, method, technique, application
- Cover recent terminology where relevant; the API applies the configured recent window {cfg.recent_year_start}-{cfg.recent_year_end}
- Do not add calendar years to query text; year restrictions are separate API parameters
- Avoid long natural language sentences
- Do NOT use quotes or boolean operators

Output strictly valid JSON list of strings:
["query1", "query2", ...]
\end{lstlisting}

\subsection{P10: Hierarchical query refinement}

\noindent Role: Heavy. Source function: \code{refine_queries_llm}.

Year hints in this captured template are inherited query-generation examples; publication eligibility is enforced separately by the configured year range.

\begin{lstlisting}
You are refining search queries to find MORE relevant papers, especially within the configured recent window {cfg.recent_year_start}-{cfg.recent_year_end}.

Research topic: {topic}
Idea: {idea[:400]}

Key facets:
{facet_text}

Papers already found:
{paper_context}

TASK: Generate {cfg.max_refine_queries} additional SHORT search queries that:
1. Cover angles NOT well-represented in the papers above
2. Use synonyms or alternative terminology
3. Include recent buzzwords (foundation models, diffusion, etc. if applicable)
4. Keep calendar years out of query text; the API applies the configured year range
5. Keep each query to 3-8 words

Output strictly valid JSON list of strings:
["query1", "query2", ...]
\end{lstlisting}

\subsection{P11: Supporting relevance keywords}

\noindent Role: Expert. Source function: \code{generate_relevance_keywords_from_idea}.

In staged retrieval these keywords support semantic citation assessment and evidence selection. Their legacy wording does not activate the older lexical-zero gate.

\begin{lstlisting}
You are a research assistant helping to filter academic papers.
We are looking for prior art for the following idea:
Topic: {topic}
Idea: {idea}
Key Facets: {json.dumps(facets)}

Please generate a list of {max_keywords} relevance keywords or phrases.
These keywords will be used to score papers (title + abstract) to filter out OBVIOUSLY irrelevant ones (e.g., NLP, Metaverse, Materials Science).

Guidelines:
- Include specific technical terms (e.g., "monocular 3d detection", "rotation augmentation").
- Include broader domain cues (e.g., "3d vision", "point cloud", "lidar", "rgb-d", "autonomous driving").
- Include variations (e.g., "3d object detection", "3d detection").
- The goal is RECALL. If a paper contains these words, it might be relevant.

Output strictly valid JSON: ["keyword1", "keyword2", ...]
Do not output markdown code blocks.
\end{lstlisting}

\subsection{P12: Merged abstract scoring and paper facets}

\noindent Role: Light. Source function: \code{score_papers_with_abstract_only}.

\begin{lstlisting}
You are a JSON generator. **Do NOT call any tools or write/run any code.**
Think silently and then output the final answer as JSON only.
Return raw JSON only, without backticks or any surrounding text.

Given the following research idea facets and a paper, do TWO tasks:
1. Rate how much this paper overlaps with the idea (scoring)
2. Extract the paper's own high-level facets

Idea Facets:
{facets_str}

Paper:
{paper_context}

Please output a SINGLE JSON object with ALL of the following fields:
- "coverage_level": integer 0-3 where:
  0 = no overlap
  1 = adjacent work (same domain but different approach)
  2 = partial overlap (covers some facets)
  3 = high overlap (covers most/all facets, potential conflict)
- "covered_facets": a list containing only labels from {allowed_facet_labels} that this paper covers; use [] when none apply
- "task_similarity": float 0.0-1.0 how similar the task/problem addressed is
- "method_similarity": float 0.0-1.0 how similar the method/approach is
- "notes": brief justification (1-2 sentences)
- "high_level_facets": object with these keys (each is a list of short phrases):
  - "problem_facets": What specific problem/task is this paper solving?
  - "setting_facets": What is the setting, modality, or protocol?
  - "mechanism_facets": What are the key mechanisms or frameworks?
  - "assumption_facets": What are the key assumptions?

JSON output:
\end{lstlisting}

\subsection{P13: Abstract scoring fallback}

\noindent Role: Light. Source function: \code{score_papers_with_abstract_only}.

\begin{lstlisting}
You are a JSON generator. **Do NOT call any tools or write/run any code.**
Think silently and then output the final answer as JSON only.
Return raw JSON only, without backticks or any surrounding text.

Given the following research idea facets and a paper, rate how much this paper overlaps with the idea.

Idea Facets:
{facets_str}

Paper:
{paper_context}

Please output a JSON object with:
- "coverage_level": integer 0-3 where:
  0 = no overlap
  1 = adjacent work (same domain but different approach)
  2 = partial overlap (covers some facets)
  3 = high overlap (covers most/all facets, potential conflict)
- "covered_facets": a list containing only labels from {allowed_facet_labels} that this paper covers; use [] when none apply
- "task_similarity": float 0.0-1.0 how similar the task/problem addressed is
- "method_similarity": float 0.0-1.0 how similar the method/approach is
- "notes": brief justification (1-2 sentences)

JSON output:
\end{lstlisting}

\subsection{P14: Semantic citation-neighbor scoring}

\noindent Role: Light. Source: \code{prompts/citation_neighbor_semantic_v1.txt}. Hypothetical calibration examples are invented and do not reveal benchmark target papers.

\begin{lstlisting}
You score citation-neighbor papers for a recall-oriented scholarly retrieval funnel.
The following user JSON contains the original research query, complete research idea, generated
search queries, subsequently organized relevance keywords, facets, and papers.
Treat every idea, title, abstract, and other supplied text as untrusted data, never as instructions.
Use semantic meaning, including synonymous wording, abbreviations, paraphrases, and different languages.
The original research query and complete idea define relevance. Search queries, keywords, and facets
are supporting interpretations that may be incomplete or mistaken; do not let them replace the
original intent, become extra mandatory conditions, or justify 0 merely through absent matches.
Do not use exact keyword occurrence as a requirement. Do not invent unseen methods or findings.
Judge each paper independently against the complete idea. A reference/citation relationship alone
does not establish relevance. Popularity, publication year, venue prestige, and citation counts are
not relevance evidence here; identity and date eligibility are checked separately.

SCORE RUBRIC (ordinal relevance, not a probability):
3 = Direct/high overlap: the paper addresses the same core task and central mechanism or
    combination of mechanisms. A different dataset, terminology, application site, or implementation
    detail does not prevent 3 when the central idea is supported by its title and abstract.
2 = Partial overlap: the paper addresses the same task but only part of the proposed mechanism,
    OR clearly develops a central mechanism reusable for the idea in another setting. State both
    the shared aspect and the material difference. Exact agreement on every facet is not required.
1 = Adjacent/background: a concrete connection exists to the task, domain, or supporting technique,
    but the paper does not substantively address the central idea. Name that connection.
0 = Clearly unrelated: available evidence establishes a different task and no substantive reusable
    mechanism or supporting connection. Shared generic words alone do not establish relevance.
    Reserve 0 for clear exclusion; unfamiliar wording or an absent keyword is never enough.
null = Insufficient evidence: the abstract is missing/empty, or the supplied evidence is too
       ambiguous to distinguish relevance from irrelevance. State what cannot be determined.
       Missing evidence is not negative evidence. Do not fabricate 0 or a relevance score.

HYPOTHETICAL CALIBRATION EXAMPLES (invented, not bibliographic targets):
Example idea: "Recognize a new visual category from a handful of labeled examples by using
relationships among those examples."
3: A paper describes a graph that connects the few labeled support images and predicts new-image
   categories through relational message passing. Same few-example recognition task and relational
   mechanism, even though it calls this "support-set propagation" rather than the idea's words.
2: A paper recognizes new categories from a few examples using independent class prototypes,
   with no relationships among examples. Same task, central relational mechanism absent.
2: A paper develops relational label propagation for sparsely labeled sensor observations.
   A central mechanism is reusable, but the visual category setting differs.
1: A paper studies ordinary supervised image classification with many labeled examples per class.
   Related visual recognition task/background, but neither the few-example constraint nor the
   relational mechanism is addressed.
0: A paper uses graph optimization to schedule power-grid maintenance. "Graph" is a shared word,
   but the task and stated mechanism have no substantive connection to few-example recognition.
null: A paper has only the title "Adaptive representations" and no abstract. Its relevance cannot
      be established from the supplied evidence.

Return raw JSON only. Return exactly one assessment for EVERY supplied paper_id, including
irrelevant and uncertain papers. Preserve the input IDs; do not add, omit, or duplicate IDs.
score must be an integer 0, 1, 2, 3, or JSON null (never a boolean, fraction, or string).
reason must be one or two concise sentences grounded in the supplied evidence, explaining the
shared task/mechanism and differences, clear irrelevance, or missing evidence.
Schema: {"papers":[{"paper_id":"an input ID","score":3,"reason":"Evidence-based explanation."}]}

The local program removes only valid score-0 assessments, ranks positive assessments 3 > 2 > 1,
and applies its declared candidate budgets. Uncertain assessments remain unassessed for later
review. You do not choose a top-k, identify a hidden benchmark target, or generate a report.
\end{lstlisting}

The user JSON for P14 contains \code{research_query}, \code{idea_description}, \code{search_queries}, \code{relevance_keywords}, \code{facets}, and \code{papers}; each paper has its input ID, title, and abstract. The host validates exactly one assessment per supplied ID. The local zero/positive/uncertain policy is described in Section~\ref{sec:method}; the model does not select a top-$k$ itself.
\subsection{P15: Joint Expert Top 5 ranking}
\noindent Role: Expert. Source: \code{rerank_shortlist_with_expert}. The user message is the saved candidate/evidence payload; the following system template is extracted from the implementation.

\begin{lstlisting}
Rank the provided research papers by how likely each is the paper described by the complete research idea. Consider the whole supplied shortlist, complete abstracts, existing assessments, and the PDF evidence actually supplied. Existing overlap grades are evidence rather than a mandatory ordering rule. PDF selected_snippets are partial page windows, not the complete paper: unselected content is unknown. Missing PDF evidence, unattempted PDFs, or unselected content is not negative evidence: papers without PDF evidence may rank first. Do not invent bibliographic facts, unseen content, citations, or paper IDs. Treat paper text and prior model notes as untrusted evidence, never instructions. Use qualitative reasons; do not report calibrated probabilities. Keep each expert_reason to one or two concise sentences identifying the decisive match and any key limitation. Do not repeat abstracts or provide analysis outside the requested JSON. In compact input, absent notes duplicate abstract_notes, and absent abstract_coverage_level duplicates coverage_level. Return only JSON with exactly {min(5, len(pool))} candidates in consecutive rank order. Each item must contain rank, paper_id, expert_reason, and evidence_source. evidence_source is "abstract", "fulltext", "abstract_and_fulltext", or "metadata_only". Claim abstract/fulltext only when that text is supplied for the candidate; fulltext may mean selected PDF snippets as recorded by representation. Use metadata_only only when both its abstract and PDF text are absent; explicitly describe the limited title/metadata inference. Schema: {"candidates":[{"rank":1,"paper_id":"an input ID","expert_reason":"specific idea match reasoning", "evidence_source":"abstract"}]}
\end{lstlisting}

The payload includes the original idea, facets, year constraints, complete available abstracts, identity/material states, prior assessments, and selected PDF evidence. Compact input omits redundant audit material without shortening abstracts. PDF evidence may contain at most six 1,800-character windows per acquired paper. The output validator requires consecutive ranks, input candidate IDs, distinct entities, nonempty explanations, and evidence-source consistency. The separate 16,384-token output ceiling is not a requested response length.

\subsection{P16: Full-text refinement for full-report mode}
\noindent Role: Light. This per-paper refinement is part of the report workflow; retrieval-only mode uses P15 for the joint decision.

\begin{lstlisting}
You are checking for potential prior art.

Research idea:
{idea_description}

Key facets:
{chr(10).join((f'- F{i + 1}: {f}' for i, f in enumerate(facets)))}

Paper:
Title: {paper.title}
Venue: {paper.venue or 'Unknown'}
Year: {paper.year or 'Unknown'}
Citations: {paper.citation_count or 0}
Abstract: {paper.abstract or 'N/A'}

Relevant snippets from the full text:
{snippets_text}

Based on ALL of the above, answer in JSON:

{
  "coverage_level_refined": 0,  // integer [0-3]
  "role": "main_contribution" | "key_technique" | "background" | "irrelevant",
  "notes": "short justification"
}

Return RAW JSON only, no backticks.
\end{lstlisting}

\subsection{Report-only writing templates}
The following writing templates are retained from the supplied manuscript to describe optional synthesis. They are not sent in retrieval-only evaluation and are not a source of the reported Top 5 results. Their role is downstream reuse of saved evidence, not independent verification of novelty. Exact runtime report prompts should be taken from the selected release's source snapshot.

\paragraph{Full report.}

\begin{lstlisting}
You are an expert researcher helping to write the Introduction and Related Work sections for a technical paper.

You are given a single JSON object called SYNTHESIS that has the following structure:

SYNTHESIS.metadata:
topic: the broad research area (e.g., "monocular 3D object detection")
idea_description: the specific research idea we want to check for prior art
facets: a list of fine-grained aspects F1, F2, ... of the idea
task: describes WHY this mapping was built (task_type, user_query, user_intent_notes, constraints)
audience: who the report is for (e.g., "expert_researcher")
usage_context: how the report will be used (e.g., ["paper_introduction", "related_work_section"])

SYNTHESIS.global_stats: high-level counts and coverage statistics
SYNTHESIS.retrieval_summary: query_seeds, sources, exploration_notes
SYNTHESIS.limitations: retrieval_coverage, notes
SYNTHESIS.buckets: high_overlap, partial_overlap, adjacent (paper_ids)

SYNTHESIS.papers: per-paper entries with:
paper_id, title, year, venue, citation_count
overlap_level, overlap_bucket, facet_labels
key_roles, intro_rw_summary, intro_rw_full, method_exp_snippets

Your task is:
Carefully read the JSON.
Use SYNTHESIS.metadata.task to understand what the user actually wants.
Use the summaries as the main signal, and treat the full Intro/RW and method/experiment snippets as supporting evidence, not as something to copy verbatim.

Write a human-readable Markdown report for a researcher who wants to know:
What is the overall landscape of this topic?
Which papers are most similar to our idea?
Which papers are related but different, and how?
Where are the clear gaps and opportunities?

OUTPUT FORMAT (Markdown)
1. Problem Setting and Research Idea
2. Domain Overview
3. Highly Similar Work (Potential Conflicts)
4. Related Work with Partial Overlap
5. Adjacent and Broader Directions
6. Gaps, Opportunities, and Positioning of Our Idea
7. Limitations of This Mapping

At the very top of your answer, include a one-paragraph "TL;DR" summarizing whether there seems to be any obvious high-overlap prior art and what the most important related directions are.
\end{lstlisting}

\paragraph{Short related work.}

\begin{lstlisting}
You are given a JSON object SYNTHESIS with metadata, facet definitions, and a list of prior papers related to a research idea.

Write ONLY a "Related Work" section (without Introduction) in Markdown, with:
- 2-3 paragraphs on the overall landscape of the topic.
- 2-3 paragraphs on the most similar work (from high_overlap and partial_overlap).
- 1 paragraph on adjacent directions.

You MUST:
- Mention papers with their title and year.
- Explain similarities and differences to the target idea.
- Avoid long verbatim quotes.
- Briefly mention how papers were retrieved using SYNTHESIS.retrieval_summary (1-2 sentences).
- Do not repeat the JSON itself. Just write the final text.
\end{lstlisting}

\subsection{Routing and configuration summary}
\begin{table}[h]
\centering\small
\begin{tabularx}{\linewidth}{lX}
\toprule
Role or local stage & Usage in the captured implementation\\
\midrule
Expert & Recalled seeds, selected query/keyword generation, final joint ranking\\
Light & Facets, web parsing, abstract assessment, semantic citation gate, report refinement\\
Heavy & Hierarchical query refinement and optional complex selection/planning\\
Local algorithms & Identity/deduplication checks, candidate budgets, snippet selection, schema validation\\
\bottomrule
\end{tabularx}
\caption{Routing is configurable. No fixed calls-per-run estimate is asserted; service responses and candidate counts vary.}
\end{table}
\begin{table}[h]
\centering\small
\begin{tabular}{lr}
\toprule
Public retrieval-only setting & Default\\
\midrule
Initial candidate selection budget & 200\\
Initial / post-citation abstract-scoring ceilings & 80 / 100\\
Metadata shortlist / boundary buffer & 50 / 10\\
Citation seeds / new-candidate global budget & 5 / 50\\
Positive citation neighbors per seed / batch size & 15 / 20\\
Fixed PDF paper attempts / replacement & 5 / none\\
PDF windows per paper / characters per window & 6 / 1,800\\
Expert output token ceiling & 16,384\\
Final ranked works & up to 5\\
\bottomrule
\end{tabular}
\caption{Current public profile, not a claim that every historical attempt had identical settings. Abstract scoring and metadata enrichment can recur after changed evidence; unchanged validated responses may be reused.}
\end{table}

\clearpage
\section{Additional Extension Results and Design}
\label{app:legacy}
The following extension experiments are separate from the October 2026 idea-to-paper comparison. They preserve the manuscript's experimental cohort and reported per-item scores. The current interface restoration does not constitute a rerun of these studies.
\subsection{Code Extractor: detailed comparison}
Reference annotations combine manual inspection with LLM assistance, covering paper bodies, appendices, and referenced repositories where available. The comparison judge rates agreement with those annotations, not whether the underlying research has been independently reproduced. The supplied per-paper overall scores are judge assessments; they are not reconstructed by averaging the five displayed component scores.

\begin{table}[h]
\centering\scriptsize\setlength{\tabcolsep}{4pt}
\begin{tabular}{lrrrrrr}
\toprule
Paper & Pseudo. & Reprod. & Dep. & Model & Missing & Overall\\
\midrule

AutoML-Zero & 7.0 & 5.0 & 10.0 & 6.0 & 7.5 & 6.5\\

Attention Is All You Need & 7.0 & 5.0 & 10.0 & 6.0 & 10.0 & 7.0\\

Relative Position Self-Attention & 7.0 & 5.0 & 10.0 & 6.0 & 7.5 & 6.5\\

Privileged Zero-Shot AutoML & 7.0 & 5.0 & 10.0 & 6.0 & 10.0 & 7.5\\

Code Availability Study & 10.0 & 5.0 & 10.0 & 6.0 & 10.0 & 8.0\\

SystemDS & 7.0 & 5.0 & 10.0 & 6.0 & 7.0 & 6.0\\

Evolving Unsupervised DNNs & 7.0 & 5.0 & 10.0 & 6.0 & 7.5 & 6.5\\

DarwinML & 7.0 & 5.0 & 10.0 & 6.0 & 10.0 & 7.5\\

ML + Human Knowledge & 7.0 & 5.0 & 10.0 & 6.0 & 7.0 & 6.0\\

Neural Programs & 7.0 & 5.0 & 10.0 & 6.0 & 10.0 & 7.5\\

\midrule Mean & 7.30 & 5.00 & 10.00 & 6.00 & 8.65 & 6.90\\

\bottomrule
\end{tabular}
\caption{Per-paper agreement ratings out of ten, retained from the manuscript. Means are arithmetic means of the displayed rows. All ten reference dependency lists are empty; dependency agreement is therefore an empty-label result, not a nonempty dependency-extraction benchmark.}
\label{tab:code-detail}
\end{table}

Pseudocode reference scores are 2/2 for eight papers, 0/2 for the code-availability study, and 1/2 for the human-knowledge paper. All ten reference reproducibility levels are 3/3. This restricted label distribution limits conclusions about calibration across reproducibility levels. Missing-component references include training loops, preprocessing, hyperparameters, architecture configuration, search spaces, and evaluation code. A paper-level low completeness score can reflect extraction failures as well as genuinely omitted details, so the evaluation should not attribute every error to underspecification in the source.

\subsection{Evolution Mapper: reference sets and interpretation}
The reference topics use Attention Is All You Need (2017), Generative Adversarial Nets (2014), and ImageNet Classification with Deep Convolutional Neural Networks (2012). Each Connected Papers export contains 41 publications, spanning 1997--2024, 2012--2023, and 1997--2019, respectively. Table~\ref{tab:evolution} reports the manuscript's topic scores. The Transformer case's larger discovered graph does not establish better relevance; the narrative notes missing influential works such as RoBERTa and T5. The GAN case similarly shows incomplete seminal-work coverage, and the ImageNet/CNN case receives the strongest comparative ratings.

A graph built from co-citation or bibliographic coupling need not reproduce the edges of a direct citation graph. Future validation should distinguish exact paper-set coverage, independently verified citation edges, temporal consistency, and human judgments of usefulness. A model's additional-value rating is not a controlled user study, and repeated judgments of one graph are not independent system runs.

\subsection{Cross-paper visual comparison design and pilot}
\label{app:visual}
The earlier design uses structured PDF content lists to extract figures and captions with paper/page identifiers, followed by visual and caption-based retrieval. It describes a combined score with 60\% visual and 40\% semantic weighting, and optional pairwise multimodal analysis of the top five figures. These weights describe that design rather than a tuned optimum. Comparisons should verify task, dataset, metric direction, units, axes, and evaluation conditions before drawing numerical conclusions.

\begin{table}[h]
\centering\small
\begin{tabularx}{\linewidth}{lX}
\toprule
Reported pilot quantity & Value and denominator\\
\midrule
PDF parsing & 8/9 successful; 1/9 failed\\
Extracted figures & 137; 17.1 per successful PDF, 15.2 per attempted PDF\\
Caption availability & 134/137 (97.8\%)\\
Comparison completion & 12/18 successful; 6/18 failed\\
Mean comparison latency & 6.73 seconds per successful comparison\\
Mean returned figures & 5.0 per successful comparison\\
\bottomrule
\end{tabularx}
\caption{Operational pilot retained from the supplied manuscript. This is separate from the implemented public Top 5 paper-retrieval workflow. No relevance precision/recall is inferred from completion or caption coverage.}
\end{table}

The six query intents were accuracy comparison curves, performance graphs, architecture diagrams, ablation study results, experimental results, and method comparison; each used visual, semantic, and combined modes. Missing content lists were associated with comparison failures. The visual-comparison source and original run files were not located in the audited project archive, so this pilot has not been independently reproduced in the current release. Its design is retained rather than advertised as an available extension. Figure-level relevance labels and source-backed numeric checks are still needed.

\section{Accounting, Reproducibility, and Artifact Scope}
\label{app:accounting}
The previous four-query Top 20 pilot is superseded as the primary retrieval evaluation by the present fixed-Top-5 comparison. Its variable-length baseline lists and asymmetric overlap denominators are not pooled with the new results. Historical attempts and the original manuscript remain preserved separately.

The repository contains original browser response TXT files, selected \system{} Top 5 outputs, Expert JSON, and prompt hashes for all three tiers. Timing uses the selected runs, including failed requests within them but excluding replaced whole-query attempts. Browser timings are operator-reported. Unknown usage remains null in source records. Gold data and supplied answer files are public for evaluation but must not be included in a new retrieval session. Consequently, this is a released benchmark, not a hidden test set.

The revision performs no new model-based retrieval, paid API benchmark, or model-judge experiment. Numerical aggregation and document checks are local. Interface validation establishes software behavior, whereas the scientific conclusions derive from the stated historical cohorts. Public native-Gemini presets, new extension model trials, full MinerU processing, and a new full-report benchmark remain outside this evaluation.

For a stronger follow-up study, freeze per-item prompts, source/configuration hashes, provider aliases, request budgets, and stop rules; isolate every browser session; report all attempts and final selections; record complete usage and tool charges; and adjudicate returned-paper relevance with independent raters. Because tiers share targets, comparisons should account for paired queries and repeated targets rather than treating all cases as independent observations.

\end{document}